# Materials discovery and properties prediction in thermal transport via materials informatics: a mini-review

Xiao Wan[1,2#], Wentao Feng[2#], Yunpeng Wang[1,2], Chengcheng Deng[2]*, Nuo Yang[1,2]*

1. State Key Laboratory of Coal Combustion, Huazhong University of Science and Technology, Wuhan 430074, China.
2. School of Energy and Power Engineering, Huazhong University of Science and Technology, Wuhan 430074, China.

# X.W. and W.F. contributed equally to this work.
* Corresponding email: dengcc@hust.edu.cn (C.D.); nuo@hust.edu.cn (N.Y.)

# Abstract

There has been an increasing demand for materials with special thermal properties, whereas experimental discovery is high-cost and time-consuming. The emerging discipline 'Materials Informatics' is an effective approach that can accelerate materials development by combining material science and big data technique. Recently materials informatics has been applied to the design of novel materials such as thermal interface materials for heat-dissipation, and thermoelectric materials for power generation. This mini-review summarized the research progress on the applications of materials informatics for the thermal transport properties prediction and discovery of materials with special thermal properties, including optimal thermal conductivity, interfacial thermal conductance and thermoelectricity efficiency. In addition, some perspectives are given for the outlook of materials informatics in the field of thermal transport.

# 1. Introduction

Thermal transport properties, such as thermal conductivity and interfacial thermal conductance (ITC) etc., play critical roles in micro/nano-electronics and opto-electronics, thermoelectrics, and some other thermal/phonon engineering areas. [1, 2] For example, there is an increasing demand for materials with high thermal conductivity that can dissipate massive heat in electronic devices. [3-6] Besides, ITC dominates the thermal dissipation of compounds with interfaces in micro/nanoscale. [7, 8] Therefore, the effective discovery of materials with high thermal conductivity or ITC is crucial to improve the performance and lifetime of a wide variety of related devices. On the other hand, thermoelectric power generation is essential for utilizing low-grade wasted heat. Researchers have been seeking materials with high conversion efficiency for decades to improve its performance, [9-14] where materials of low thermal conductivity are essential for this purpose.

Due to limitations of cost, time and hardware, the discovery of materials with adaptive thermal properties remains challenges in both experiments and simulations. [15] The materials informatics (Fig. 1) opens up a brand new way for accelerating the discovery of materials with special properties. [16, 17] It is an emerging area of materials science, [16-18] which is based on simulations or experiments in materials science and machine-learning algorithms. [16] The machine-learning algorithms are critical for materials informatics to decipher processing-structure-property-performance relationships in materials via the results of simulation or experiment. Moreover, that will help to accelerate property prediction and materials discovery. In the past, materials informatics has been successfully applied in the search for the materials or structures with special thermal transport properties, such as thermal conductivity and ITC. [19-22] Considering there are some studies in this brand new opened field, it is necessary to have a review on the progress and give outlook on future work, which would be of helpful to the development of materials informatics in thermal transport.

In this paper, a brief review is given for the recent investigations on the applications of materials informatics in thermal transport. Firstly, we make a brief introduce of materials informatics. Then, the studies applying materials informatics are summarized in the following areas respectively: thermal conductivity, interfacial thermal conductance and thermoelectric conversion efficiency. Lastly, it is proposed that remaining challenges and outlook for future investigations on this research topic.

## 2. Materials informatics

The framework of materials informatics mainly consists of three parts: Data procurement, acquiring data generated by simulations or experiments in materials science; Data representation, systematic storage of representative structure and property information about these materials; Data mining, data analysis aimed at searching for novel materials, predicting properties or gaining new physical insights. [17] The specific contents of the three steps are described as follows, respectively.

### 2.1 Data procurement

Data procurement is to acquire physical properties and structural information of given materials. Both calculations (such as first-principles, [20, 23] molecular dynamics [22, 24, 25] and lattice dynamics [26, 27] etc.), experiments [28] and online libraries [29], have been used to collect these data. With above different techniques, the database repositories containing effective training data can be constructed validly.

### 2.2 Data representation

Data representation means a systematic storage of representative structure and property information. The pivotal part of the data representation is the selection of the characters (e.g., formation energies, band structure, density of states, magnetic moments) to describe the materials, which are called 'descriptors'. The descriptors represent different kind of materials and they are a part of the input in data mining. One purpose of materials informatics is to establish the mapping relations between descriptors and target properties, which are thermal transport properties here. Thus, good descriptors are the key of effective materials informatics. Once a series of good descriptors are identified, the search for optimum materials or properties prediction within the database can be performed intrinsically or extrinsically. [30]

## 2.3 Data mining

Data mining aims at searching for novel materials or exploring new physical insights, where machine learning is widely used. [17] The machine-learning algorithms mainly used in materials informatics are the aspect of supervised learning, whose task is finding a function that maps an input to an output based on samples. [31] Through the training models built by learning algorithms, they could select or predict materials with novel properties easily. Some popular algorithms are Bayesian optimization, random-forest regression, and artificial neural network. A brief introduction of these algorithms are given in the following.

Bayesian optimization is a well-established technique for global optimization of black-box functions. [32, 33] Bayesian prediction models, most commonly based on Gaussian process, are usually employed to predict the black-box function, where the uncertainty of the predicted function is also evaluated as predictive variance. [34] The Bayesian optimization algorithm (BOA) typically works by assuming the unknown predicted function was sampled from a Gaussian process and maintains a posterior distribution for this function as observations are made. [33] The procedure for Bayesian optimization is as follows. First, a Gaussian process model is developed from two randomly selected observations taken from the database. The model is updated by (i) sampling the point at which the observation property is expected to be the best and (ii) updating the model including the observation at the sampled point. These two steps are repeated until all the data are sampled. [30]

Random forests [35] is an prominent ensemble method adapted from bagging,which combines multiple decision trees into one predictive model to improve performance. [19] It is relatively robust to various problems, such as compound classification, and can handle outliers data or high-dimensional data well. [35, 36] A random forest model consist of K decision trees can be established in three steps. Firstly, generate K sets of data from the initial dataset by boostrap method. Secondly, grow a tree with the particular random selection algorithm and get the predictions for each data point. Thirdly, final prediction is done by weighted vote (in classification) or weighted average (in regression) of the whole forest predictions. [36] Besides performing the prediction task, random forest also provides an intrinsic metric to evaluate the importance of each descriptor. [19]

Artificial neural networks (ANNs) and deep neural networks are well-developed machine learning methods that mimic human brains to learn the relationships between certain inputs and outputs from experience. [37] The ANN has been successfully applied in fields of modeling and prediction in many thermal engineering systems recent years. [38-40] It has become increasingly attractive in the last decade. The assets of ANN compared to classical methods are its high speed and simplicity, which decrease engineering effort. [28, 41, 42] The most basic and commonly used ANN consists of at least three or more layers, an input layer, an output layer, and a number of hidden layers. [28] The number of neurons in the input layer equals the number of parameters in the material selection process. The output layer represents the fitness of the candidate materials. In addition, the hidden layer represents relationships between the input and output layers. Through training and testing stages, the ANN model can be well established. In the training stage, the network is trained to predict an output based on input data. The training stage is stopped when the testing error is within the tolerance limits. In the testing stage, the network is tested to stop or continue training it according to measures of error. [28, 39-41]

## 3. Thermal conductivity

As a hot topic, the thermal conductivity is one of the most important properties of materials. In some cases, the materials with ultralow or superhigh thermal conductivity are essential for many applications. [2, 3, 8] In search of compounds with ultralow or superhigh thermal conductivity, several works have been done on lattice thermal conductivity by materials informatics. Apart from discovery of new materials, there are also some works about thermal conductivity prediction model on liquids build by machine-learning algorithms.

In studying lattice thermal conductivity (LTC, $\kappa_\omega$), via random-forest regression among 79000 entries of the database (Fig. 2), Carrete *et al*. proposed three half-Heusler semiconductors, whose LTCs are below $5\ Wm^{-1}K^{-1}$, for further experimental study. [19] They also found the material with larger average atomic radius in positions A and B tend to have lower thermal conductivity. More important, the efficient methods are introduced for reliably estimating the $\kappa_\omega$ for a series of compounds, which are based on a combination of random-forest regression and first-principles calculations. That is, there is a very good prospect of machine-learning methods for applications in accelerating material design. In this study, we should notice that the performance in predicting LTC using machine-learning algorithms is largely affected by selecting descriptors. To find good descriptors, in 2017, Tanaka's group proposed a procedure to generate a series of compound descriptors from simple atomic representations. [30] When it was applied to the LTC data set, these descriptors in terms of Bayesian optimization exhibit good predictive performance, which verified the accuracy of the approach. Apart from the bulk lattice thermal conductivity, machine-learning algorithms also can predict thermal conductivity of composite materials. In August 2018, Wei *et al.* proposed the models obtained from the support vector regression, Gaussian process regression and convolution neural network. [43] The prediction of effective thermal conductivity based on these models and effective medium theory matches well with experimental data.

Besides predicting thermal conductivity of solids, there are some works focusing on fluids. In early 2009, Kurt *et al.* reported an artificial neural network (ANN) model to predict the thermal conductivity of ethylene glycol-water solutions based on experimentally measured variables. [28] The regression analysis between the predicted by the model and the experimental values proved the high accuracy of the ANN model. The superiority of this model compared to practical experiments lies in less time and cost, which is exactly the advantage of machine-learning algorithms. In addition, a multilayer perceptron-artificial neural network (MLP-ANN) model was reported by Zendehboudi *et al.* in order to predict the effective thermal conductivity of nanofluids with desired accuracy. [44]

In the above, it is introduced that the materials informatics has successfully applied to discover materials with novel lattice thermal conductivity. In the materials informatics, different descriptors should be investigated and compared in order to decrease the deviation of the prediction. Moreover, the prediction models built by machine-learning algorithms can also predict the thermal conductivity of composite materials and liquids.

## 4. Interfacial thermal transport

The interfacial thermal transport plays an important role in the thermal management of high power micro- and opto-electronic devices where a large number of interfaces exist. [7, 8, 45] Prediction of interfacial thermal transport property is important for guiding the discovery of interfaces with very low or very high thermal boundary resistance (TBR), which can further adjust the thermal conductivity of the whole system.

In 2017, three different machine-learning algorithms were used to predicted TBR by Zhan *et al.*, and they also compared their results with the commonly-used acoustic mismatch model (AMM) and diffuse mismatch model (DMM) to verify the accuracy. [46] Three different machine-learning algorithms included generalized linear regression (GLR), Gaussian process regression (GPR) and support vector regression (SVR). The correlation coefficient ($R$) (Fig. 3) showed the correlation between experimental values and the values predicted by different methods, which demonstrated greater accuracy of these methods compared to traditional AMM and DMM. [46] One year later, via trained machine-learning models, Yang *et al.* predicted the ITC between graphene and hexagonal boron-nitride (h-BN) with only the knowledge of system temperature, coupling strength and tensile strains. [22] The machine-learning algorithms used in these predictions were linear regression, polynomial regression, decision trees, random forests and artificial neural networks. In addition, the performance of these different methods were compared with molecular dynamics simulation. It was shown that the artificial neural networks performed best in the predictions. These results illustrate the simplicity and accuracy of machine-learning methods for the prediction of interfacial thermal transport properties.

In addition to the prediction of interfacial thermal transport property, the optimization of interfacial structures is also significant for the discovery of materials with special thermal transport properties. In order to minimize or maximize the interfacial thermal conductance (ITC) across Si-Si and Si-Ge interfaces (Fig. 4), Ju. *et al.* proposed a method combining atomistic Green's function(AGF) and Bayesian optimization in May

2017, which could obtain the optimal interfacial structures with a few calculations. [21] Then, they applied it to Si/Ge superlattice and figured out the interface with highest and lowest ITC by calculating a few interface structure. These results deepen the understanding of the mechanisms in interfacial thermal transport. It also indicates the cost-effectiveness of materials informatics in designing nanostructures with special thermal transport properties.

As for interfacial thermal transport properties, the thermal boundary resistance was predicted via three different machine-learning algorithms. Moreover, the accuracy was proved by comparing them with experimental data. Then, methods to accelerate the discovery of interfacial structures with highest or lowest interfacial thermal conductance were proposed.

## 5. Thermoelectricity

The performance of thermoelectric materials is ranked by the dimensionless thermoelectric figures of merit (ZT), which is defined as $T\sigma S^2/\kappa$, where $T$, $\sigma$, $S$, and $\kappa$ are temperature, electrical conductivity, Seebeck coefficient, and thermal conductivity, respectively. [47, 48] A good thermoelectric material with a high ZT should possess a low thermal conductivity, high electrical conductivity and Seebeck coefficient. In despite of the fact that these three properties are closely correlated, the materials informatics has been used in the design and search for high-ZT thermoelectric materials without exhaustive experiments and simulations while some works major in the prediction of thermoelectric properties of materials.

In 2014, Carrete *et al.* used the decision tree methods to unveil the rules that judge the thermoelectric performance of a nanograined half-Heusler compound good or bad. [49] They found two key properties for high ZT, which are a large lattice parameter and either a wide gap (at high temperatures), or a large effective mass of holes (at room temperature). These results could stimulate the experimental research for improving thermoelectric performance of half-Heuslers semiconductors. Furthermore, Tanaka's group combined the Bayesian optimization and first-principles anharmonic lattice-dynamics calculations, in order to find materials with ultralow thermal conductivity. [20] They discovered 221 materials with very low thermal conductivity in the library containing 54,779 compounds in 2015. Two of them (Fig. 5) even have an electronic band gap < 1 eV, which makes them promising for thermoelectric applications. Compared to other methods, this strategy doesn't need much computation cost owing to less initial data. However, Tanaka's methods could just figure out the materials with low thermal conductivity instead of high ZT. In 2018, Yamawaki *et al.* realized the goal of predicting ZT. They used Bayesian optimization to get an optimally designed graphene structure with higher ZT. [50] The procedure is quite the same with their group's previous work. [21] Bayesian optimization showed its strength to accelerate the searching procedure.

Compared to the feasibility, the practicability of these methods is also important. To give some advice on the selection of materials for experimental researchers, the recommendation engine (http://thermoelectrics.citrination.com) based on machine learning was proposed by Gaultois *et al.* in 2016. [29] To ensure the accuracy, they tested an example set of compounds generated by the engine, $RE_{12}Co_5Bi$ (RE = Gd, Er), which exhibits surprising thermoelectric performance (Fig. 6). The materials predicted by this engine with low thermal conductivity and high electrical conductivity were also confirmed experimentally. It is suggested that this paradigm could promote the discovery of good thermoelectric materials greatly. In a word, the research mentioned above will greatly accelerate the procedure of finding materials with high ZT.

Besides thermal conductivity, Seeback coefficient is another important factor to determine ZT. In order to predict Seeback coefficient of different materials precisely, Furmanchuk *et al.* used random forests algorithm and got a great success. [51] The prediction was quite accurate, which means that we don't need to try hard to synthesis or calculate materials and get its Seeback coefficient. In addition to the prediction of thermoelectricity properties, materials informatics can also be used to optimize structures for enhancing the thermoelectric figure of merit. Apart from searching for materials with high ZT, predictions of thermoelectric properties are also quite important. In order to predict Seeback coefficient of different materials precisely, Furmanchuk *et al.* used random forests algorithm and got a great success. [51] The prediction was quite accurate, which means that we don't need to try hard to synthesis or calculate materials and get its Seeback coefficient.

Apart from these studies, the high-throughput calculations have already been applied successfully in the discovery of new thermoelectric materials. [29, 52-58]

In this part, the machine learning-based methods were proposed in search of materials with high ZT and prediction of thermoelectric properties. Especially, some rules determining the ZT of materials were revealed by the machine-learning algorithm. Moreover, an open machine learning-based recommendation engine was proposed for experimental researchers with an aim to find new materials with high ZT.

## 6. Summary and perspectives

To date, there have been more and more investigations on the interdisciplinary field of materials informatics and thermal science. In this mini-review, we have summarized recent representative results on thermal transport by materials informatics. The procedures of practical implementation of materials informatics were presented, especially several important machine-learning algorithms were descript. A comprehensive framework and main conclusions were exhibited in discovering materials with optimal thermal conductivity, interfacial thermal conductance and thermoelectricity efficiency. Moreover, several critical factors affecting the discovery efficiency and predictive efficacy were discussed. It is also emphasized that the superiority of materials informatics in discovering novel materials for thermal transport.

Even though plenty of achievements have been reached via materials informatics, some critical elements still need to be put in place. When carrying out the materials informatics, it is still challenging to make efficient machine-learning algorithm codes, pick up fast and effective descriptors, and transfer data to practical knowledge or physical pictures. Besides, the main challenge lies in the physical interpretation of the process by machine learning. The underlying physical mechanism cannot be fully understood by machine learning only, which need the help of other theoretical or simulative methods. Advances in studying the heat transfer in nanomaterials/nanostructures are needed by machine learning. When making data preparation for nanomaterials/nanostructures, especially for complex structures, it takes much time for calculations due to a larger simulation cell. So, the time for data preparation is challenging.

# Acknowledgements

The work was sponsored by National Natural Science Foundation of China No. 51576076 (N.Y.), No. 51606072 (C.D.), No. 51711540031 (N.Y. and C.D.), Natural Science Foundation of Hubei Province No. 2017CFA046 (N.Y.) and Fundamental Research Funds for the Central Universities No. 2016YXZD006 (N.Y.). We are grateful to Xiaoxiang Yu, Dengke Ma and Han Meng for useful discussions. The authors thank the National Supercomputing Center in Tianjin (NSCC-TJ) and China Scientific Computing Grid (ScGrid) for providing assistance in computations.

# Figure

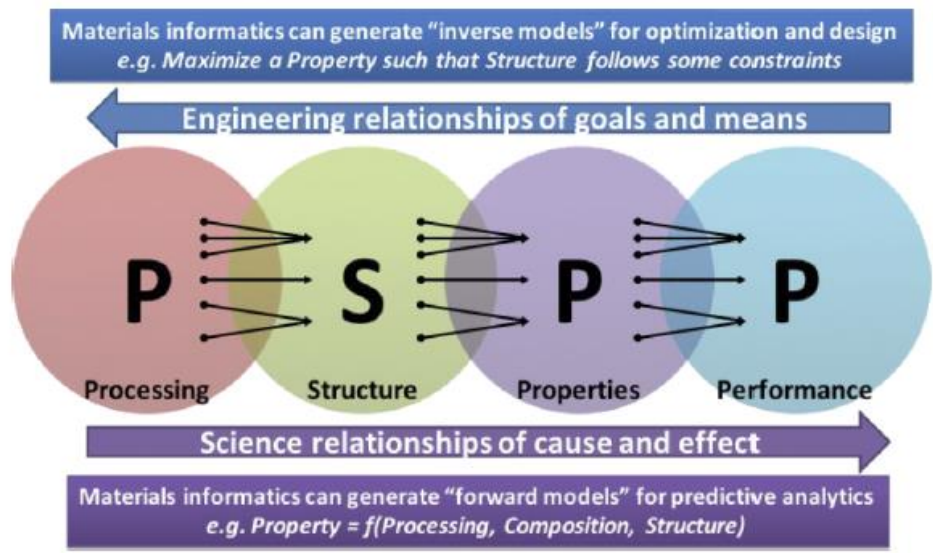

Fig. 1. The processing-structure-property-performance relationships of materials science and engineering, and how materials informatics approaches can help decipher these relationships via forward and inverse models. [16]

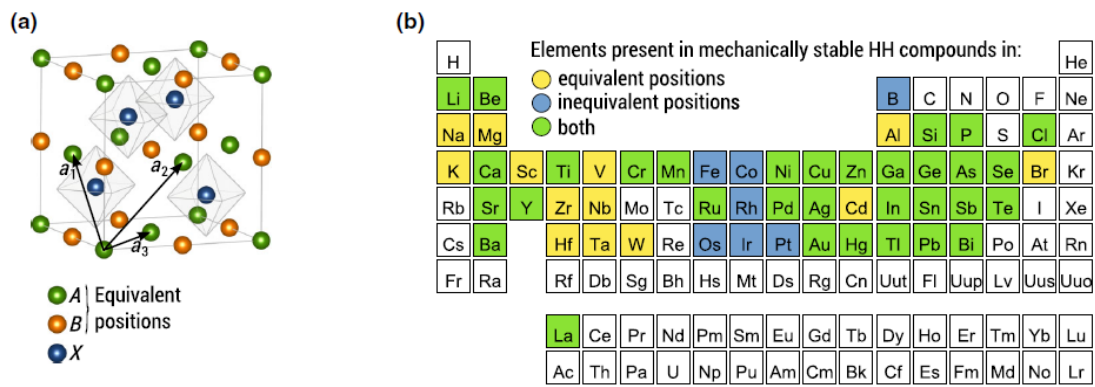

Fig. 2. (a) Prototype Half-Heusler structure with primitive vectors and a conventional cell. (b) Elements considered in this study. [19]

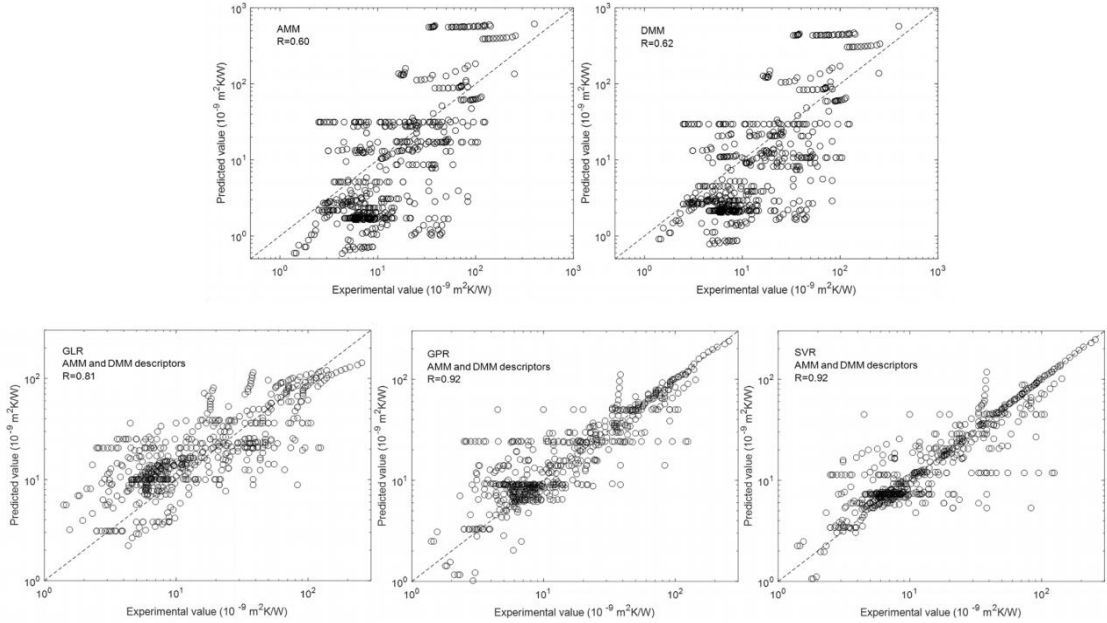

Fig. 3. Correlation between the experimental values and the values predicted by the AMM, DMM, GLR, GPR, and SVR using the same descriptors. [46]

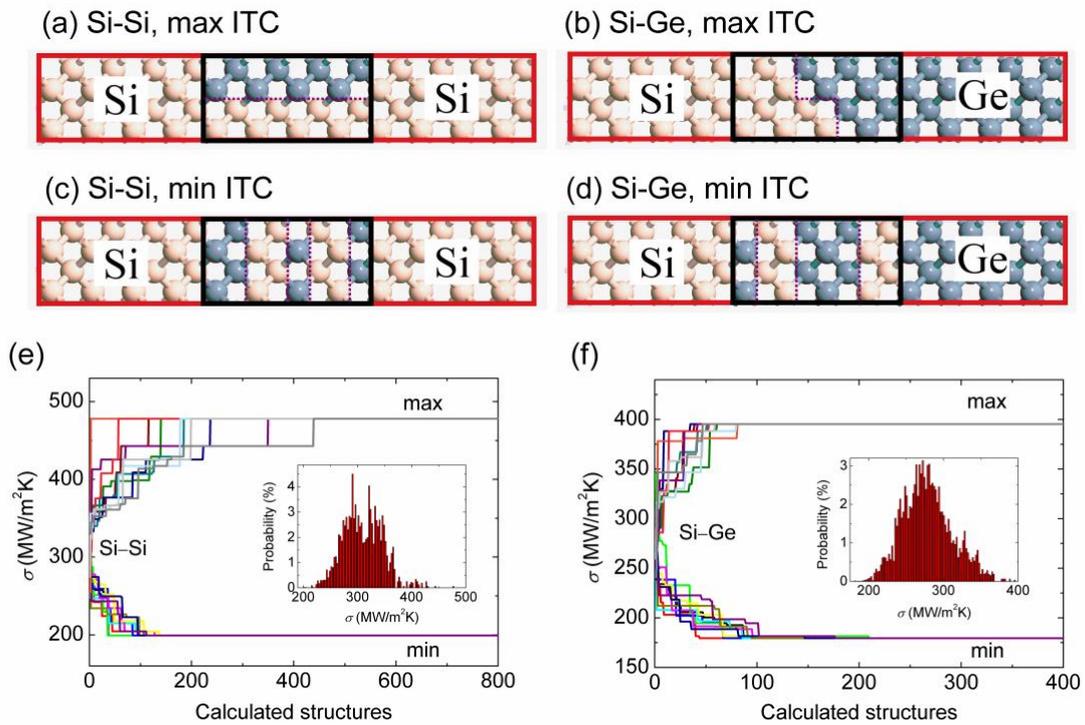

Fig. 4. Interfacial Si/Ge alloy structure optimization (a)-(d) Optimal structures with the maximum and minimum interfacial thermal conductance (ITC) for Si-Si and Si-Ge interface. (e), (f) The 10 optimization runs with different initial choices of candidates, where the insets show the probability distributions of ITC obtained from calculations of all candidates. [21]

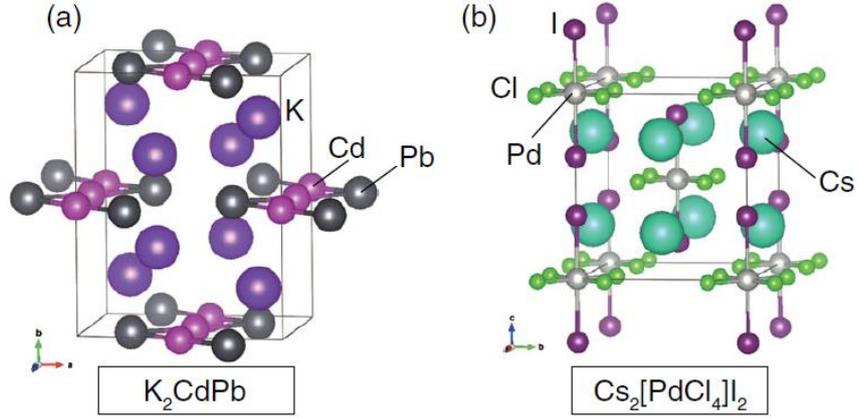

Fig. 5. Crystal structures of $K_2CdPb$ and $Cs_2[PdCl_4]I_2$ predicted to show the low LTC of <0.5 W/mK (at 300 K) and narrow band gap of < 1 eV. [20]

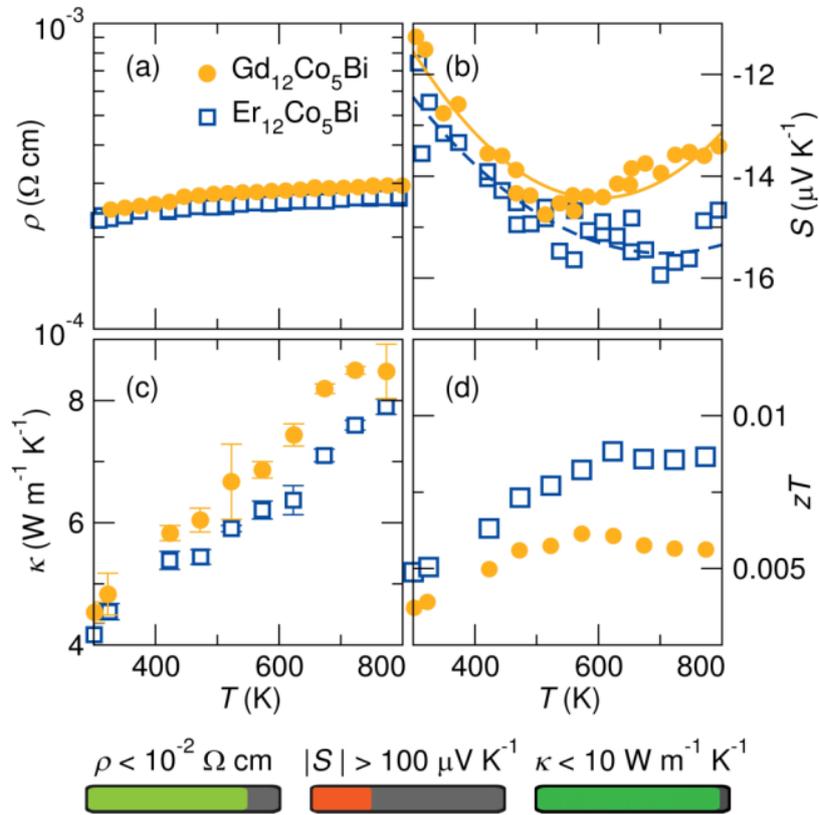

Fig. 6. Thermoelectric characterization of $RE_{12}Co_5Bi$ (RE = Gd, Er). (a) Electrical resistivity, (b) Seebeck coefficient, (c) thermal conductivity, and (d) thermoelectric figure of merit $zT$ as a function of temperature. [29]